\documentclass[prb,twocolumn,superscriptaddress,amsmath,amssymb,showpacs]{revtex4}

\usepackage{graphicx}
\usepackage{dcolumn}
\usepackage{bm}

\begin{document}


\title{Dynamical control of electron spin coherence in a quantum dot}

\author{Wenxian Zhang}
\affiliation{Ames Laboratory, Iowa State University, Ames, Iowa
50011, USA}

\author{V. V. Dobrovitski}
\affiliation{Ames Laboratory, Iowa State University, Ames, Iowa
50011, USA}

\author{Lea F. Santos}
\affiliation{Department of Physics and Astronomy, Dartmouth
College, Hanover, New Hampshire 03755, USA}

\author{Lorenza Viola}
\affiliation{Department of Physics and Astronomy, Dartmouth
College, Hanover, New Hampshire 03755, USA}

\author{B. N. Harmon}
\affiliation{Ames Laboratory, Iowa State University, Ames, Iowa
50011, USA}

\date{\today}

\begin{abstract}
We investigate the performance of dynamical decoupling methods at
suppressing electron spin decoherence from a low-temperature nuclear
spin reservoir in a quantum dot.  The controlled dynamics is studied
through exact numerical simulation, with emphasis on realistic pulse
delays and long-time limit.  Our results show that optimal performance
for this system is attained by a periodic protocol exploiting
concatenated design, with control rates substantially slower than
expected from the upper spectral cutoff of the bath.  For a known
initial electron spin state, coherence can saturate at long times,
signaling the creation of a stable ``spin-locked'' decoherence-free
subspace.  Analytical insight on saturation is obtained for a simple
echo protocol, in good agreement with numerical results.
\end{abstract}


\maketitle

Developing and benchmarking strategies for decoherence suppression
in spin nanosystems are of vital importance to various areas of
quantum physics, from quantum control theory to quantum device
technologies. A central spin 1/2 interacting with a bath of $N$
external spins provides a natural testbed for detailed
analysis~\cite{CSP}. This conceptually simple system shows a rich
variety of decoherence regimes, paving the way to the
understanding of more complex scenarios, such as decoherence of
many-spin central systems. A prominent example is an electron
localized in a quantum dot (QD) at experimentally relevant
sub-Kelvin temperatures and moderate (sub-Tesla) magnetic fields,
where the hyperfine coupling with a bath of nuclear spins is the
dominant decoherence channel.  Although electron spins in QDs have
a broad range of potential applications in spintronics
\cite{Zutic04} and scalable quantum information processing
\cite{Loss98}, the coherence time $T_2$ is very short, $T_2
\gtrsim T_2^*$, where the free induction decay (FID) time $T_2^*
\sim 10$~ns in a typical GaAs QD~\cite{dqde05}.  While suggestive
proposals exist, to increase $T_2$ by achieving high bath spin
polarization or bath disentanglement, or by narrowing the nuclear
spin distribution~\cite{Burkard99},
methods viable in a wider range of physical parameters are actively
sought.

The long correlation time and distinctively non-Markovian behavior of
the nuclear spin reservoir make the electron spin an ideal candidate
for pulsed spin resonance \cite{NMRDDBook} and dynamical decoupling
(DD) techniques \cite{ViolaDD}. In double-QD devices, for instance,
spin singlet refocusing has been experimentally
demonstrated~\cite{dqde05}. For a single QD in a large external
magnetic field, $B_0 \gtrsim 1$~Tesla, where the nuclei simply dephase
the electron spin, Hahn spin echoes and their
Carr-Purcell-Meiboom-Gill modifications are expected to enhance
$T_2^*$ by at least an order of magnitude in GaAs
QDs~\cite{Hahnecho05,Taylor06}. For weaker bias fields, the nuclear
spin coupling induces both dephasing and relaxation, and the use of
higher-level DD schemes has been invoked recently
\cite{Khodjasteh05,Bergli}.  However, the formal limits of
applicability of these analyses are very restrictive and extremely
hard to meet in practice.

In this paper, we perform a quantitative study of the electron spin
decoherence DD problem in regimes which are important for experimental
DD implementations in QDs, yet have received little attention so
far. Focusing on the challenging situation of zero external field,
where dephasing and relaxation must be simultaneously eliminated, we
investigate to what extent the very stringent formal limitations of DD
methods may be relaxed. Using exact numerical simulations, we identify
promising DD protocols, and show that for pulse delays {\em up to a
factor $\sqrt{N}$ longer} than naively expected from analytical
bounds, they are still capable of extending the coherence time by 2-3
orders of magnitude. Special emphasis is devoted to the asymptotic
long-time limit, where error accumulation is crucial and neither
intuition based on Magnus expansion (ME) nor the quasi-static
approximation is reliable {\it a priori}.  Provided that the initial
electron spin state is known, nearly perferct coherence preservation
may be achieved for indefinitely long times.  Such a {\em saturation}
is related to the creation of a {\it stable\/} decoherenc free
subspace (DFS)~\cite{Viola00}, and may be exploited for stabilizing
the electron spin polarization in a QD.


{\em Model and DD Setting.} The dynamics of a single electron spin $S$
coupled to a bath $B$ of $N$ nuclear spins is described by a total
Hamiltonian of the form $H = H_S + H_{SB} + H_B$, where $H_S=H_0 S_z$
is the electron Zeeman energy in an external magnetic field $B_0$,
$H_{SB} =\sum_{k=1}^N A_k {\bf S} \cdot {\bf I}_k$ the hyperfine
contact interaction between the electron spin and the nuclei, and
$H_B= \sum_{k>l}^N \Gamma_{kl} ({\bf I}_k \cdot {\bf I}_l-3I^z_k
I^z_l)$ the intrabath dipolar coupling between nuclear spins
~\cite{Al-Hassanieh06,Dobrovitski03,Zhang06,Taylor06}. ${\bf S}$ and
${\bf I}_k$ denote the electron and the $k$-th bath spin operators,
respectively. The nuclear spin value is set to
$I_k=1/2$~\cite{foot1}. We focus on the limit of zero external field
$B_0=0$ and assume that the bath is initially unpolarized.  The FID
time scale is $T_2^* = (NA^2/8)^{-1/2}$, where $A=(\sum_k
A_k^2/N)^{1/2} \approx 10^{-4}\mu$eV for typical GaAs QDs with
$N=10^6$ \cite{Zhang06}. Time is measured in units of $1/A$.

Under ideal control assumptions, DD is implemented by subjecting the
electron spin to sequences of instantaneous $\pi_{\hat{\bf n}}$
rotations along appropriate control axes $\hat{\bf n}$, equally
separated by the interval $\tau$. A variety of DD protocols exist,
based on both deterministic \cite{ViolaDD,Khodjasteh05} and randomized
\cite{Viola-random,closed} design. In {\em cyclic} DD, the control
propagator is steered through a DD group of unitary operations ${\cal
G}=\{g_j\}$, $j=0, 1, \ldots, |{\cal G}|-1$, in a predetermined order,
as opposed to {\em randomized} DD where the future control path is not
known in advance.  Changing $g_i$ to $g_j$ requires the application of
a DD pulse $P_{i,j}=g_{j}g_{i}^{\dagger}$.  Thanks to the existence of
a periodicity time scale $T_c = |{\cal G}|\tau$, the analysis of
cyclic DD has been mostly carried out within average Hamiltonian
theory \cite{NMRDDBook}, upper performance bounds being determined by
the dominant non-zero corrections in the ME for the time evolution
operator.  Average Hamiltonian theory no longer applies to randomized
DD, where the evolution is most directly studied in a logical frame
that follows the applied control \cite{Viola-random}.

Periodic DD (PDD) is the simplest {\em non-selective} cyclic protocol,
ensuring that the unwanted evolution is removed to first order in the
ME at every $T_n=nT_c$, $n \in {\mathbb N}$, in the short $T_c$
limit. For a single spin, PDD is based on the irreducible Pauli group
${\cal G}_P=\{I,X,Y,Z \}$ \cite{ViolaDD}, which requires two-axis
control sequences of the form $C_1=C_0XC_0ZC_0XC_0Z$, $C_0$ denoting a
free evolution period.  Improvement over PDD may be gained by
symmetrized and/or concatenated design. Symmetric DD (SDD) guarantees
that all odd terms in the ME are cancelled, with $T_c$ twice as long
as PDD.  Concatenated DD~\cite{Khodjasteh05} relies on a temporal
recursive structure, so that at level $\ell+1$ the protocol is
$C_{\ell+1}=C_{\ell}XC_{\ell}ZC_{\ell}XC_{\ell}Z$. Here, we {\em
truncate} the concatenation procedure at a certain level and repeat a
periodic sequence, referred to as PCDD, after every $4^{\ell}\tau$
(e.g., $\ell=2$ leads to PCDD$_2$). As representatives among
stochastic protocols, we consider naive random DD (NRD), which
corresponds to uniformly random pulses over ${\cal G}$, and symmetric
random path DD (SRPD), where a path to traverse ${\cal G}$ is chosen
at random and then symmetrized as in SDD \cite{closed}.

The use of control pulses may suit two purposes: (i) complete
decoupling of the system from the bath, so that electron spin
coherence is enhanced for an {\em arbitrary} initial state; (ii)
preservation of a {\em specific} initial state, in which case the DD
sequence may be tailored accordingly. Two performance metrics are then
appropriate. For a fixed initial state $|\psi\rangle$, we use the
input-output fidelity $F(T)={\rm Tr}[\rho_S (T)\rho_S (0)]$, where
$\rho_S(T)$ is the reduced density operator of $S$ at time $T$
starting from $|\psi\rangle$ and tracing out the bath.  For an unknown
initial state, we invoke minimum pure-state fidelity
$F_m(T)=\mbox{min}_{|\psi\rangle} F(T)$.  Analytical bounds on the
expected fidelity decay for various DD protocols have been obtained
for short evolution times~\cite{Viola-random,closed,Khodjasteh05},
which calls for numerical analysis in the long-time
regime. Simulations also make it possible to explore DD performance
for values of $T_c$ beyond the strict convergence domain of the ME,
$\omega_c T_c \ll 1$, where the highest frequency component
$\omega_c\approx \sum_k |A_k|/4 \sim N A/4$. Let $\sigma$ denote the
power spectrum width of the environmental coupling, $2\sigma \approx
(\sum_k A_k^2)^{1/2}= \sqrt{N} A$ \cite{Melikidze04}. We shall
consider $\tau \sim 1/2\sigma$, thus $T_c\ge 4\tau \sim
\sqrt{N}\omega_c^{-1}$. To solve the time-dependent Schr\"odinger
equation of the entire $S$ plus $B$ system, we apply the Chebyshev
polynomial expansion method to the evolution operator
\cite{Dobrovitski03}, and choose $A_k >0$ as uniformly random numbers.

\begin{figure}[bt]
\includegraphics[width=3in]{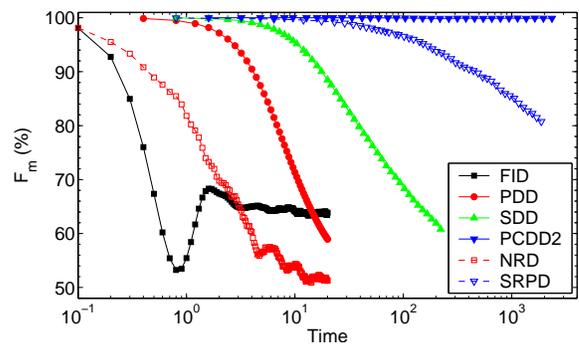}
\caption{(Color online) Minimum fidelity vs. time in the logical
frame with $\tau=0.1$.  Hamiltonian parameters are $H_0=0$,
$\Gamma_0=0$ and $N=15$.
For deterministic DD, data points are acquired at the completion of
each cycle, while for NRD and FID this is done after every $\tau$, and
for SRPD after every $8\tau$. Random protocols are averaged over
$10^2$ control realizations.} \label{fig:cp}
\end{figure}

{\em Unknown initial state.} In Fig.~\ref{fig:cp} we compare $F_m(T)$
for the above-mentioned DD protocols.  Because the characteristic time
scale $\tau_D$ for nuclear dipolar dynamics due to $H_B$ is (at least)
two orders of magnitude slower than the one due to $H_{SB}$ in typical
QDs, setting $H_B=0$ is justified for practically relevant time
regimes. All schemes lead to substantial enhancement of the electron
spin coherence, PCDD$_2$ showing the most dramatic
improvement. Although, for this system both SDD and PCDD$_2$ remove
$H_{SB}$ to second order in the ME, the higher performance of PCDD$_2$
reflects its superiority in reducing coherent error accumulation. The
poor performance of NRD is expected, since its advantages over
deterministic DD emerge only when ${\cal G}$ is large. Contrary to the
case of closed systems~\cite{closed}, SRPD does not match PCDD$_2$ in
the relevant parameter range, confirming the fact that irreducible DD
groups and slow baths are predicted to be especially favorable for
concatenated control~\cite{Khodjasteh05}.

\begin{figure}[t]
\includegraphics[width=3in]{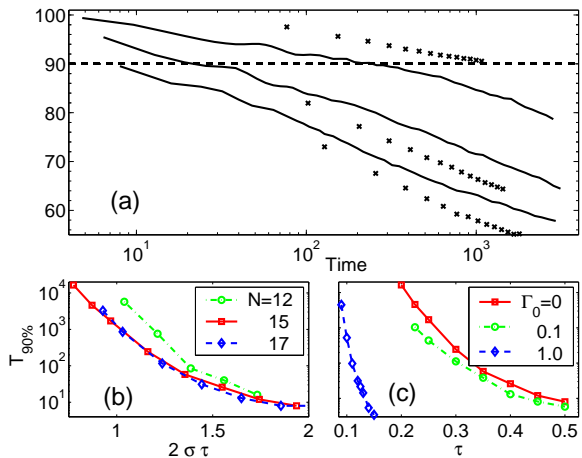}
\caption{(Color online) (a) PCDD$_2$ (solid lines) and PCDD$_4$
(crosses) for $\tau=0.3$, $0.4$, $0.5$, top to bottom. (b) and
(c): $T_{90\%}$ vs. $\tau$ for PCDD$_2$; different bath sizes (b),
intrabath interactions (c).  $H_0=0$ in all panels, $N=15$ in (a)
and (c).} \label{fig:cdd}
\end{figure}

Motivated by the above results, we proceed with a more in-depth
analysis of the PCDD protocol. Fig.~\ref{fig:cdd}(a) compares the
performance of two levels of concatenation, $\ell=2,4$, for different
values of $\tau$. As expected, the results deteriorate as $\tau$
increases but, interestingly, PCDD$_4$ becomes worse than
PCDD$_2$. Also interesting is the exponential fidelity decay of both
protocols at long times.  Fig.~\ref{fig:cdd}(b) illustrates, for each
value of $2 \sigma \tau$ and different $N$, the instant of time
$T_{90\%}$ where $F_m(T)$ for PCDD$_2$ reaches 90\%. The results are
reasonably close to each other, particularly for larger $N$,
supporting their applicability up to realistic situations with $N\sim
10^6$.  Lastly, we analyze the effect of $H_B$, which becomes
important once the coherence time is longer than $\tau_D$. Let
$\Gamma_{kl}$ be uniformly random numbers in $[-\Gamma_0,
\Gamma_0]$. To avoid demanding long time simulations, we increase
$\Gamma_0$ manually up to values comparable to $A_k$.  The results are
shown in Fig.~\ref{fig:cdd}(c), where a two dimensional $3 \times 5$
QD with nearest neighbor intrabath coupling is considered. PCDD$_2$
performance is significantly affected by a bath with fast
dynamics. Although such a regime is not directly relevant to standard
GaAs QDs, further investigation of randomized DD is necessary whenever
$H_B$ and $H_{SB}$ compete.

\begin{figure}[t]
\includegraphics[width=2.8in]{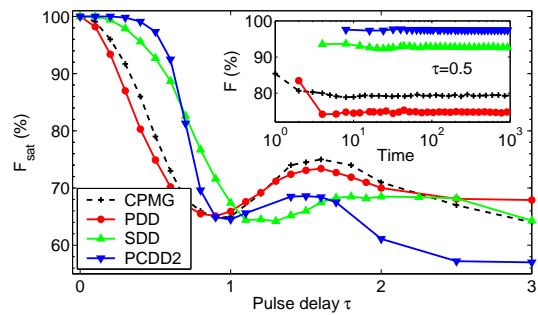}
\caption{(Color online) Fidelity saturation for CPMG, PDD, SDD,
and PCDD$_2$ starting from an initial state along the half cycle
direction. Hamiltonian parameters as in Fig.~\ref{fig:cp}. The
inset shows how the asymptotic value $F_{sat}$ is reached.  In the
main panel, a number of pulses sufficient to reach saturation and
close the cycle of each protocol is chosen, $n_p \sim 50$.  }
\label{fig:ddfs}
\end{figure}

{\em Known initial state.} If the electron spin is initially pointing
along a known direction, {\em cyclic} DD protocols able to stabilize
the input-output fidelity value for extremely long times may be used.
This is shown in Fig.~\ref{fig:ddfs} (inset), where the curves $F(T)$
plateau after the application of a sufficient number $n_p$ of pulses.
While asymptotic saturation behavior has been reported for purely
dephasing spin-boson models with {\em arbitrary} initial spin states
\cite{ViolaDD,saturation-old}, the directional dependence observed
here reflects the lack of a preferred direction in the error process
generated by $H_{SB}$: a preferred direction only emerges through the
``effective field'' {\em created} by the control sequence, and
long-time stability depends on proper alignment between such effective
field and the initial state.  In magnetic resonance language, the
resulting saturation effect is closely related to the ``pedestals'' of
the long-time magnetization signal in pulsed spin-locking
experiments~\cite{Lock}.  From a control standpoint,
it indicates the dynamical generation of a stable one-dimensional DFS
via DD~\cite{Viola00}.  Consider first a {\em selective} echo
protocol, say a single-axis PDD along the $z$ direction, ${\cal
G}_Z=\{ I, Z\}$, with a corresponding (asymmetric) pulse sequence
$C_Z=C_0ZC_0Z$ -- which we refer to as CPMG.  For sufficiently small
$\tau$, symmetrization is enforced along the $z$ axis, as described by
a lowest-order Hamiltonian commuting with ${\cal G}_Z$ in the ME, and
a corresponding effective field along $z$~\cite{ViolaDD,Lock}: initial
$S_z$-eigenstates are (approximate) eigenstates of the decoupled
evolution, whereas components perpendicular to the DFS are lost in the
long-time regime.  For non-selective DD protocols based on the
irreducible group ${\cal G}_P$, all directions are approximately
preserved for short times due to maximal averaging, yet long-time
stability again occurs along the direction of the dominant term in the
ME.  Notice that the latter also coincides with the {\em half cycle}
direction of the sequence: e.g., PDD may be obtained from
concatenation of two CPMGs, $C_1=C_X Y C_X Y= C_Y \circ C_X$,
identifying the outer $y$ direction as the stable one.  Similarly, for
SDD and PCDD$_2$, the saturated components are $z$ and $y$,
respectively.

Quantitative results on the dependence of the saturation value
upon control parameters are given in Fig.~\ref{fig:ddfs}.
Interestingly, a crossing between SDD and PCDD$_2$ occurs at
$\tau\sim 0.8$. In particular, the CPMG sequence, which is not a
maximal DD scheme for the Hamiltonian in question, leads to
saturation values comparable to the other protocols, thus it may
be useful in settings where accurate control along two axes may
not be available. Moreover, its simplicity allows for a direct
analytical study of the saturation effect.  Within the quasistatic
bath approximation (QSA)~\cite{Taylor06, Zhang06}, let $A_k=A$,
${\mathbf I}=\sum_k {\mathbf I}_k$, and $M=I_z$. After $n$ CPMG
cycles, the survival probability of the initial state
$|\Psi(0)\rangle= |\uparrow\rangle \otimes |I,M\rangle$ is given
by $ |\langle \Psi(0)|\Psi(2n\tau)\rangle|^2=1- {(C^2/ B^2)}
\tan^2\theta \cos^2 2n\theta$, where $C=A\sqrt{(I-M)(I+M+1)}$,
$B=A(M+1/2)$, $\tan\theta=d/\sqrt{1-d^2}$, $d =
-(B/\Omega)\sin(\Omega \tau/2)$, and $\Omega^2 = B^2+C^2$.
Averaging over the nuclear spin bath and taking the limit of large
$n$ and $N$, $F \rightarrow F_{sat} = 1 - (1/2)\int dI dM P(I,M)$
$ (C^2/ B^2) \tan^2\theta$ with $P(I,M)\simeq (I /D\sqrt{2\pi
D}\;)e^{-I^2/2D}$ and $D=N/4$ for an unpolarized bath
\cite{Melikidze04}. In the limit of small $\tau$, we obtain
$F_{sat} = 1- (1/16) \tau^2 A^2 N = 1-\tau^2/2T_2^{*2}$. For
randomly distributed $A_k$, $A^2N \mapsto \sum_k A_k^2$.

\begin{figure}[t]
\includegraphics[width=2.7in]{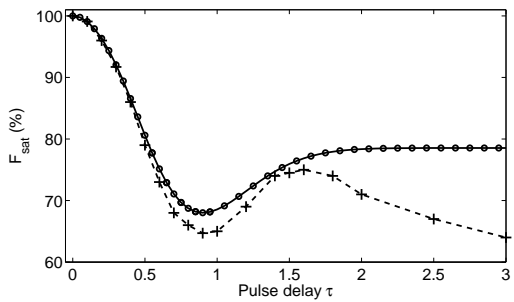}
\caption{Fidelity saturation vs. pulse delay for CPMG from a known
initial state. $H$ as in Fig.~\ref{fig:cp}. Circles - QSA results;
solid line - classical random field model; dashed line with plus
signs - exact numerical simulations.} \label{fig:CPMG}
\end{figure}

Fig.~\ref{fig:CPMG} compares the above analytical result with the
saturation value predicted by a semiclassical approximation, which
treats the nuclear Overhauser field as an effective random magnetic
field with zero average, but finite variance \cite{Merkulov}. The two
curves superimpose, consistent with the fact that for large $N$ the
Overhauser field induced by ${\bf I}$ indeed approaches a classical
field. Also shown are data from exact numerical simulations of a
quantum spin bath with randomly distributed $A_k$. Remarkably, for
short pulse delays the exact and QSA results are in good agreement, in
spite of QSA being well known to be only valid for times comparable to
$T_2^*$. Thus, (i) DD effectively extends the region of validity of
QSA; and (ii) saturation is entered before the QSA becomes invalid,
allowing QSA to accurately predict $F_{sat}$ for short $\tau$.

In summary, we have quantitatively characterized DD of an electron
spin coupled to a nuclear spin bath, with emphasis on long-time
behavior. We find that DD can significantly enhance the coherence time
for an arbitrary initial state, actual performance depending on both
control and physical parameters.  For a known initial state, the
possibility of long-time saturation has been established numerically
and analytically, which may provide a way for preserving the electron
spin polarization without the need of a strong permanent magnetic
field.  While, from a practical standpoint, the estimated control time
scales ($\sim 1$ ns) are roughly an order of magnitude away from
current pulsing capabilities in GaAs QDs, experimental progress is
steady.  In particular, single-electron spin rotations have been
demonstrated both in gate-defined GaAs and self-assembled
QDs~\cite{Chen04}.  In addition, multipulse CPMG-DD has also been
realized, not only in standard NMR and ESR experiments, but also in
single solid-state centers \cite{Fraval05}, which share many relevant
features with electron spins in QDs.  These advances support the hope
that the experimental implementation
of more complex protocols will be achievable in the near future.

We thank D. G. Cory, A. Imamoglu, J. J. Longdell, and A. J. Rimberg
for discussions. This work was partially carried out at the Ames
Laboratory, which is operated for the U.S. DOE by Iowa State
University under contract No. W-7405-82 and was supported by the
Director of the Office of Science, Office of Basic Energy Research of
the U.S. DOE. Work at Ames was also supported by the NSA and ARDA
under ARO contract DAAD 19-03-1-0132. L. F. S. and L. V. also
acknowledge partial support from the NSF through Grant No.
PHY-0555417.


\begin{thebibliography}{widest-label}
\bibitem{foot1} Up to appropriate parameter renormalization, the
analysis is not qualitatively dependent upon this choice.
\end{thebibliography}
\end{document}